# NetChain: Scale-Free Sub-RTT Coordination
## (Extended Version)

Xin Jin[1], Xiaozhou Li[2], Haoyu Zhang[3], Nate Foster[2,4],
Jeongkeun Lee[2], Robert Soulé[2,5], Changhoon Kim[2], Ion Stoica[6]

[1]*Johns Hopkins University,* [2]*Barefoot Networks,* [3]*Princeton University,*
[4]*Cornell University,* [5]*Università della Svizzera italiana,* [6] *UC Berkeley*

## Abstract

Coordination services are a fundamental building block of modern cloud systems, providing critical functionalities like configuration management and distributed locking. The major challenge is to achieve low latency and high throughput while providing strong consistency and fault-tolerance. Traditional server-based solutions require multiple round-trip times (RTTs) to process a query. This paper presents NetChain, a new approach that provides scale-free sub-RTT coordination in datacenters. NetChain exploits recent advances in programmable switches to store data and process queries entirely in the network data plane. This eliminates the query processing at coordination servers and cuts the end-to-end latency to as little as half of an RTT—clients only experience processing delay from their own software stack plus network delay, which in a datacenter setting is typically much smaller. We design new protocols and algorithms based on chain replication to guarantee strong consistency and to efficiently handle switch failures. We implement a prototype with four Barefoot Tofino switches and four commodity servers. Evaluation results show that compared to traditional server-based solutions like ZooKeeper, our prototype provides orders of magnitude higher throughput and lower latency, and handles failures gracefully.

## 1 Introduction

Coordination services (e.g., Chubby [1], ZooKeeper [2] and etcd [3]) are a fundamental building block of modern cloud systems. They are used to synchronize access to shared resources in a distributed system, providing critical functionalities such as configuration management, group membership, distributed locking, and barriers. These various forms of coordination are typically implemented on top of a *key-value store* that is replicated with a consensus protocol such as Paxos [4] for *strong consistency* and *fault-tolerance*.

High-throughput and low-latency coordination is essential to support interactive and real-time distributed applications, such as fast distributed transactions and subsecond data analytics tasks. State-of-the-art in-memory transaction processing systems such as FaRM [5] and DrTM [6], which can process hundreds of millions of transactions per second with a latency of tens of microseconds, crucially depend on fast distributed locking to mediate concurrent access to data partitioned in multiple servers. Unfortunately, acquiring locks becomes a significant bottleneck which severely limits the transaction throughput [7]. This is because servers have to spend their resources on (*i*) processing locking requests and (*ii*) aborting transactions that cannot acquire all locks under high-contention workloads, which can be otherwise used to execute and commit transactions. This is one of the main factors that led to relaxing consistency semantics in many recent large-scale distributed systems [8, 9], and the recent efforts to avoid coordination by leveraging application semantics [10, 11]. While these systems are successful in achieving high throughput, unfortunately, they restrict the programming model and complicate the application development. A fast coordination service would enable high transaction throughput without any of these compromises.

Today's server-based solutions require multiple end-to-end round-trip times (RTTs) to process a query [1, 2, 3]: a client sends a request to coordination servers; the coordination servers execute a consensus protocol, which can take several RTTs; the coordination servers send a reply back to the client. Because datacenter switches provide sub-microsecond per-packet processing delay, the query latency is dominated by host delay which is tens to hundreds of microseconds for highly-optimized implementations [12]. Furthermore, as consensus protocols do not involve sophisticated computations, the workload is communication-heavy and the throughput is bottlenecked by the server IO. While state-of-the-art solutions such as NetBricks [12] can boost a server to process tens of millions of packets per second, it is still orders of magnitude slower than a switch.

We present NetChain, a new approach that leverages the power and flexibility of new-generation programmable switches to provide scale-free sub-RTT coordination. In contrast to server-based solutions, NetChain is an in-network solution that stores data and processes queries entirely within the network data plane. We stress that NetChain is not intended to provide a new theoretical answer to the consensus problem, but rather to provide a systems solution to the problem. Sub-RTT implies that NetChain is able to provide coordination within the network, and thus reduces the query latency to as little as half of an RTT. Clients only experience processing delays caused by their own software stack plus a relatively small network delay. Additionally, as merchant switch ASICs [13, 14] can process several billion packets per second (bpps), NetChain achieves orders of magnitude higher throughput, and scales out by partitioning data across multiple switches, which we refer to as *scale-free*.

The major challenge we address in this paper is building *a strongly-consistent, fault-tolerant, in-network key-value store* within the functionality and resource limit of the switch data plane. There are three aspects of our approach to address this challenge. (*i*) We leverage the switch on-chip memory to store key-value items, and process both *read* and *write* queries directly in the data plane. (*ii*) We design a variant protocol of *chain replication* [15] to ensure strong consistency of the key-value store. The protocol includes a routing protocol inspired by segment routing to correctly route queries to switches according to the chain structure, and an ordering protocol based on sequence numbers to handle out-of-order packet delivery. (*iii*) We design an algorithm for fast failover that leverages network topologies to quickly resume a chain's operation with remaining nodes, and an algorithm for failure recovery that restores the partial chains to the original fault-tolerance level and leverages virtual groups to minimize disruptions.

NetChain is incrementally deployable. The NetChain protocol is compatible with existing routing protocols and services already in the network. NetChain only needs to be deployed on a few switches to be effective, and its throughput and storage capacity can be expanded by adding more switches. In summary, we make the following contributions.

- We design NetChain, a new strongly-consistent, fault-tolerant, in-network key-value store that exploits new-generation programmable switches to provide scale-free sub-RTT coordination.
- We design protocols and algorithms tailored for programmable switches to ensure strong consistency (§4) and fault-tolerance (§5) of the key-value store.
- We implement a NetChain prototype with Barefoot Tofino switches and commodity servers (§7). Evaluation results show that compared to traditional server-based solutions like ZooKeeper, NetChain provides orders of magnitude higher throughput and lower latency, and handles failures gracefully (§8).

Recently there has been an uptake in leveraging programmable switches to improve distributed systems. NetChain builds on ideas from two pioneering works in particular: NetPaxos [16, 17] and NetCache [18]. Net-

|  | Server | Switch |
|---|---|---|
| **Example** | NetBricks [12] | Tofino [13] |
| **Packets per sec.** | 30 million | a few billion |
| **Bandwidth** | 10-100 Gbps | 6.5 Tbps |
| **Processing delay** | 10-100 $\mu s$ | $< 1 \mu s$ |

Table 1: Comparison of packet processing capabilities.

Paxos uses programmable switches to accelerate consensus protocols, but it does not offer a replicated key-value service, and the performance is bounded by the overhead of application-level replication on servers. NetCache uses programmable switches to build a load-balancing cache for key-value stores, but the cache is not replicated and involves servers for processing write queries. In comparison, NetChain builds a strongly-consistent, fault-tolerant key-value store in the network data plane. We discuss NetChain's limitations (e.g., storage size) and future work in §6, and related work in detail in §9.

## 2 Background and Motivation

### 2.1 Why a Network-Based Approach?

A network data-plane-based approach offers significant advantages on latency and throughput over traditional server-based solutions. Moreover, such an approach is made possible by the emerging programmable switches such as Barefoot Tofino [13] and Cavium XPliant [19].

**Eliminating coordination latency overhead.** The message flow for a coordination query is:

**client → coordination servers → client.**

Coordination servers execute a consensus protocol such as Paxos [4] to ensure consistency, which itself can take multiple RTTs. NOPaxos [20] uses the network to order queries and eliminates the message exchanges between coordination servers. It reduces the latency to two message delays or one RTT, which is the *lower bound* for server-based solutions. As shown in Table 1, since data-center switches only incur sub-microsecond delay, the host delay dominates the query latency. By moving coordination to the network, the message flow becomes:

**client→ network switches → client.**

Because the network time is negligible compared to host delay, a network-based solution is able to cut the query latency to one message delay or sub-RTT, which is better than the lower-bound of server-based solutions. Note that the sub-RTT latency is not a new theoretical answer to the consensus problem, but rather a systems solution that eliminates the overhead on coordination servers.

**Improving throughput.** The workload of coordination systems is communication-heavy, rather than computation-heavy. While varying in their details, consensus protocols typically involve multiple rounds of message exchanges, and in each round the nodes examine their messages and perform simple data comparisons and updates. The throughput is determined by how fast the nodes can process messages. Switches are specifically designed and deeply optimized for packet processing and switching. They provide orders of magnitude higher throughput than highly-optimized servers (Table 1). Alternative designs like offloading to NICs and leveraging specialized chips (FPGAs, NPUs or ASICs) either do not provide comparable performance to switch ASICs or are not immediately deployable due to cost and deployment complexities.

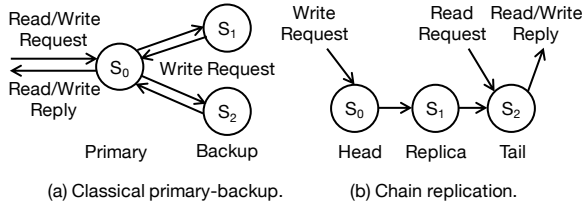

Figure 1: Primary-backup and chain replication.

## 2.2 Why Chain Replication?

Given the benefits, the next question is how to build a replicated key-value store with programmable switches. NetCache [18] has shown how to leverage the switch on-chip memory to build a key-value store on *one* switch. Conceivably, we can use the key-value component of NetCache and *replicate* the key-value store on *multiple* switches. But the challenge in doing so would be how to ensure strong consistency and fault-tolerance.

**Vertical Paxos.** We choose to realize Vertical Paxos [21] in the network to address this challenge. Vertical Paxos is a variant of the Paxos algorithm family. It divides a consensus protocol into two parts, i.e., *a steady state protocol* and *a reconfiguration protocol*. The division of labor makes it a perfect fit for a network implementation, because *the two parts can be naturally mapped to the network data and control planes*. (*i*) The steady state protocol is typically a primary-backup (PB) protocol, which handles read and write queries and ensures strong consistency. It is simple enough to be implemented in the network data plane. In addition, it only requires $f$+1 nodes to tolerate $f$ node failures, which is lower than $2f$+1 nodes required by the ordinary Paxos, due to the existence of the reconfiguration protocol. This is important as switches have limited on-chip memory for key-value storage. Hence, given the same number of switches, the system can store more items with Vertical Paxos. (*ii*) The heavy lifting for fault-tolerance is offloaded to the reconfiguration protocol, which uses an auxiliary master to handle reconfiguration operations like joining (for new nodes) and leaving (for failed nodes). The auxiliary master can be mapped to the network control plane, as modern datacenter networks already have a logically centralized controller replicated on multiple servers.

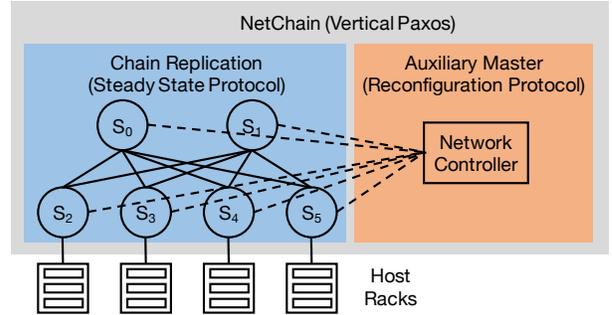

(a) NetChain architecture.

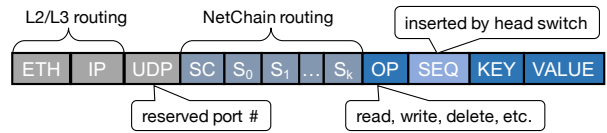

(b) NetChain packet format.

Figure 2: NetChain overview.

While it seems to move the fault-tolerance problem from the consensus protocol to the auxiliary master, Vertical Paxos is well-suited to NetChain because reconfigurations such as failures (on the order of minutes) are orders of magnitude less frequent than queries (on the order of microseconds). So handling queries and reconfigurations are mapped to data and control planes, respectively.

**Chain Replication.** We design a variant of chain replication (CR) [15] to implement the steady state protocol of Vertical Paxos. CR is a form of PB protocols. In the classical PB protocol (Figure 1(a)), all queries are sent to a primary node. The primary node needs to keep some state to track each write query to each backup node, and to retry or abort a query if it does not receive acknowledgments from all backup nodes. Keeping the state and confirming with all backup nodes are costly to implement with the limited resources and operations provided by switch ASICs. In CR (Figure 1(b)), nodes are organized in a chain structure. Read queries are handled by the tail; write queries are sent to the head, processed by each node along the chain, and replied by the tail. Write queries in CR use fewer messages than PB ($n$+1 instead of $2n$ where $n$ is the number of nodes). CR only requires each node to apply a write query locally and then forward the query. Receiving a reply from the tail is a direct indication of query completion. Thus CR is simpler than PB to be implemented in switches.

## 3 NetChain Overview

We design NetChain, an in-network coordination service that provides sub-RTT latency and high throughput. It provides a strongly-consistent, fault-tolerant key-value store abstraction to applications (Figure 2(a)).

**NetChain data plane (§4).** We design a replicated key-value store with programmable switches. Both read and write queries are *directly processed in the switch data plane* without controller involvement. We design a variant of CR to ensure strong consistency of the key-value store. Coordination queries use a custom network format based on UDP (Figure 2(b)) and the processing logic of NetChain is invoked by a reserved UDP port.

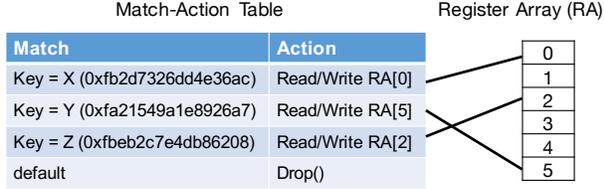

Figure 3: Key-value index and storage.

**NetChain control plane (§5).** The controller handles system reconfigurations such as switch failures, which are orders of magnitude less frequent than read and write queries. The controller is assumed to be reliable by running a consensus protocol on multiple servers, as required by Vertical Paxos. Note that NetChain can also work with traditional distributed network control planes. In such scenarios, we need to implement an auxiliary master for handling system reconfigurations, and the network control plane should expose an interface for NetChain to manage NetChain's tables and registers in switches. NetChain does not need access to any other switch state, such as forwarding tables.

**NetChain client.** NetChain exposes a key-value store API to applications. A NetChain agent runs in each host, and translates API calls from applications to queries using our custom packet format. The agent gathers returned packets from NetChain and generates API responses. The complexity of the in-network key-value store is hidden from applications by the agent.

## 4 NetChain Data Plane

**Problem statement.** The data plane provides a replicated, in-network key-value store, and handles read and write queries directly. We implement CR to guarantee strong consistency, which involves three specific problems: ($R1$) how to store and serve key-value items in each switch; ($R2$) how to route queries through the switches according to the chain structure; ($R3$) how to cope with best-effort network transport (i.e., packet reordering and loss) between chain switches.

**Properties.** NetChain provides strong consistency: ($i$) reads and writes on individual keys are executed in some sequential order, and ($ii$) the effects of successful writes are reflected in subsequent reads. NetChain assumes trustworthy components; otherwise, malicious components can destroy these consistency guarantees.

**Algorithm 1** ProcessQuery(pkt)
- *sequence*: the register array that stores sequence numbers
- *value*: the register array that stores values
- *index*: the match table that stores array locations of keys

1: $loc \leftarrow index[pkt.key]$
2: **if** $pkt.op == read$ **then**
3:     Insert value header field $pkt.val$
4:     $pkt.val \leftarrow value[loc]$
5: **else if** $pkt.op == write$ **then**
6:     **if** $isChainHead(pkt)$ **then**
7:         $sequence[loc] \leftarrow sequence[loc] + 1$
8:         $pkt.seq \leftarrow sequence[loc]$
9:         $value[loc] \leftarrow pkt.val$
10:    **else if** $pkt.seq > sequence[loc]$ **then**
11:        $sequence[loc] \leftarrow pkt.seq$
12:        $value[loc] \leftarrow pkt.val$
13:    **else** Drop()
14: Update packet header and forward

### 4.1 Data Plane Key-Value Storage

**On-chip key-value storage.** Modern programmable switch ASICs (e.g., Barefoot Tofino [13]) provide on-chip register arrays to store user-defined data that can be *read* and *modified* for each packet at line rate. NetChain separates the storage of *key* and *value* in the on-chip memory. Each *key* is stored as an entry in a match table, and each *value* is stored in a slot of a register array. The match table's output is the index (location) of the matched key. Figure 3 shows an example of the indexing. Key X is stored in slot 0 of the array, and key Z is stored in slot 2. NetChain uses the same mechanism as NetCache [18] to support variable-length values with multiple stages.

**UDP-based key-value query.** We leverage the capability of programmable switches to define a custom packet header format (Figure 2(b)) and build a UDP-based query mechanism (Algorithm 1). The core header fields for key-value operations are OP (which stands for *operator*), KEY and VALUE. Other fields are used for routing (§4.2) and ordered delivery (§4.3). NetChain supports four operations on the key-value store: Read and Write the value of a given key; Insert and Delete key-value items. Read and Write queries are entirely handled in the data plane at line rate. Delete queries invalidate key-value items in the data plane and requires the control plane for garbage collection. Insert queries require the control plane to set up entries in switch tables, and thus are slower than other operations. This is acceptable because Insert is a less frequent operation for the use cases of coordination services. Most queries are reads and writes on configuration parameters and locks that already exist. Nevertheless, if necessary in particular use cases, data plane insertions are feasible using data structures like d-left hashing, but at the cost of low utilization of the on-chip memory.

**Data partitioning with consistent hashing.** NetChain uses consistent hashing [22] to partition the key-value store over multiple switches. Keys are mapped to a hash ring, and each switch is responsible for several continuous segments on the ring. Virtual nodes [23] are used to help evenly spread the load. Given $n$ switches, NetChain maps $m$ virtual nodes to the ring and assign $m/n$ virtual nodes to each switch. Keys of each segment on the ring are assigned to $f+1$ subsequent virtual nodes. In cases where a segment is assigned to two virtual nodes that are physically located on the same switch, NetChain searches for the following virtual nodes along the ring until we find $f+1$ virtual nodes that all belong to different switches.

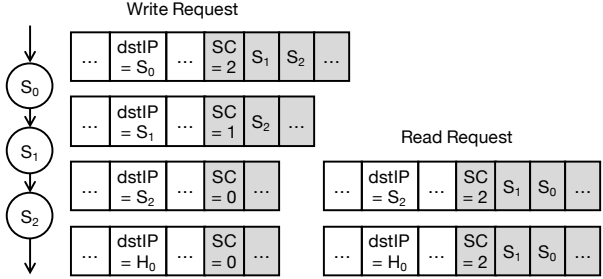

Figure 4: NetChain routing.

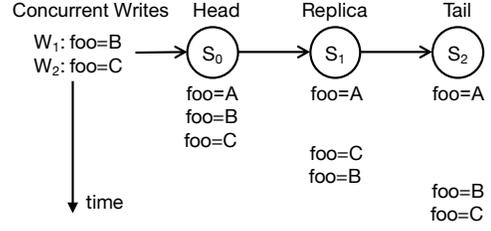

(a) Problem of out-of-order delivery.

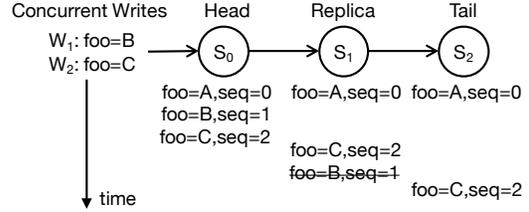

(b) Serialization with sequence numbers.

Figure 5: Serializing out-of-order delivery.

## 4.2 NetChain Routing

**NetChain routing protocol.** Our goal is to route queries for switch processing according to the chain structure. This is different from *overlay routing* that uses servers as intermediate hops and *underlay routing* that specifies hop-by-hop routing. We require that certain hops must be visited in the chain order, but do not care how queries are routed from one chain node to the next. While this is similar to *segment routing* and *loose source routing*, the write and read queries visit the chain switches in different orders, and read queries only visit the tail of the chain. We build the NetChain routing protocol on top of existing underlay routing protocols. This allows us to partially deploy NetChain with only a few switches being NetChain nodes, and take advantage of many properties provided by existing routing protocols, e.g., fast rerouting upon failures.

Specifically, we assign an IP address for each switch, and store an IP list of the chain nodes in the packet header (Figure 2(b)). *SC*, which is short for switch count, denotes the number of remaining switches in the chain. The destination IP in the IP header indicates the next chain node for the query. When a switch receives a packet and the destination IP matches its own address, the switch decodes the query and performs the read or write operation. After this, the switch updates the destination IP to the next chain node, or to the client IP if it is the tail.

Write queries store chain IP lists as the chain order from head to tail; read queries use the reverse order (switch IPs other than the tail are used for failure handling with details in §5). The chain IP lists are encoded to UDP payloads by NetChain agents. As we use consistent hashing, a NetChain agent only needs to store a small amount of data to maintain the mapping from keys to switch chains. The mapping only needs to be updated when the chain is reconfigured, which does not happen frequently. As will be described in §5, we only need to make local changes on a few switches to quickly reflect chain reconfigurations in the network, followed by a slower operation to propagate the changes to all agents.

**Example.** The example in Figure 4 illustrates NetChain routing for chain $[S_0, S_1, S_2]$. A write query is first sent to $S_0$, and contains $S_1$ and $S_2$ in its chain IP list. After processing the query, $S_0$ copies $S_1$ to the destination IP, and removes $S_1$ from the list. Then the network uses its underlay routing protocol to route the query based on the destination IP to $S_1$, unaware of the chain. After $S_1$ processes the packet, it updates the destination IP to $S_2$ and forwards the query to $S_2$. Finally, $S_2$ receives the query, modifies the query to a reply packet, and returns the reply to the client $H_0$. A read query is simpler: it is directly sent to $S_2$, which copies the value from its local store to the packet and returns the reply to the client.

## 4.3 In-Order Key-Value Update

**Problem of out-of-order delivery.** CR is originally implemented on servers and uses TCP to serialize messages. However, when we move CR to switches and use UDP, the network data plane only provides best-effort packet delivery. Packets can arrive out of order from one switch to another, which introduces consistency problems that do not exist in the server-based CR. Figure 5(a) shows the problem of out-of-order delivery. We have a key foo that is replicated on $S_0$, $S_1$ and $S_2$ with initial value A. We have two concurrent write queries, which modify the value to B and C respectively. $W_1$ and $W_2$ are reordered when they arrive at $S_1$, and are reordered again when they arrive at $S_2$. foo has inconsistent values on the three switch replicas. Clients would see foo changing from C to B if $S_2$ fails. To make things worse, if $S_1$ also fails, clients would see the value of the item reverting to C again. This violates the consistency guarantees provided by CR.

**Serialization with sequence numbers.** We use sequence numbers to serialize write queries, as shown in Algorithm 1. Each item is associated with a sequence number, which is stored in a dedicated register array that shares the same key indexes with the value array. The head switch assigns a *monotonically increasing* sequence number to each write query, and updates its local value. Other switches only perform write queries with higher sequence numbers. Figure 5(b) shows how the out-of-order delivery problem is fixed by sequence numbers. $S_1$ drops $W_1$ as $W_1$ carries a lower sequence number. The value is consistent on the three replicas.

## 4.4 Comparison with Original CR

NetChain replaces the servers used in the original CR with programmable switches, which requires new solutions for *R*1, *R*2 and *R*3 (see the preamble of §4). These solutions are limited by switch capabilities and resources. Here we describe the differences and the intuitions behind the correctness of the new protocol.

The solution to *R*1 (§4.1) implements *per-key* read and *write* queries, as opposed to *per-object* read and *update* queries. NetChain does not support multi-key transactions. The motivation for this change is due to characteristics of modern switches, which have limited computation and storage. Note that per-object operations are more general than per-key operations, as objects can include multiple keys or even the entire database. In principle it is possible to pack multiple application values into a single database value for atomic and consistent updates, although there are practical limitations on value sizes, as discussed in §6. Likewise, updates are more general than writes, as they need not be idempotent and may require multiple operations to implement. However, writes are still sufficient to implement state updates and so are often provided as primitives in the APIs of modern coordination services [2, 3].

The solution to *R*2 (§4.2) routes queries through switches, which does not change the CR protocol.

The solution to *R*3 (§4.3) delivers messages between chain switches over UDP. The original CR assumes ordered, reliable message delivery (implemented with TCP), which is not available on switches. NetChain uses *sequence numbers* to order messages, and relies on *client-side retries* (e.g., based on a timeout) when messages are lost. Because datacenter networks are typically well managed, the already-small packet loss rate can be further reduced by prioritizing NetChain packets, since coordination services are critical and do not have large bandwidth requirements. In addition, because writes are idempotent, retrying is benign. By considering every (repeated) write as a new operation, the system is able to ensure strong consistency. Note that client-side retries are also used by the original CR to handle message losses between the clients and the system, because it is difficult to track the liveness of the clients which are peripheral to the system.

CR and its extensions have been used by many key-value stores to achieve strong consistency and fault-tolerance, such as FAWN-KV [24], Flex-KV [25] and HyperDex [26]. The design of NetChain is inspired by these server-based solutions, especially FAWN-KV [24] and its Ouroboros protocol [27]. The unique contribution of NetChain is that it builds a strongly-consistent, fault-tolerant key-value store into the *network*, with the switch on-chip key-value storage, a routing protocol for chain-based query routing and an ordering protocol for query serialization, which can all be realized in the *switch data plane*. In addition, as we will show in §5, compared to the failure handling in the Ouroboros protocol [27], NetChain leverages the network topology to reduce the number of updated nodes for failover, and relies on the controller to synchronize state in failure recovery instead of directly through the storage nodes themselves.

## 4.5 Protocol Correctness

To prove that our protocol guarantees consistency in unreliable networks with switch failures, we show that the following invariant always holds, which is a relaxation of the invariant used to establish the correctness of the original CR [15].

**Invariant 1** *For any key k that is assigned to a chain of nodes $[S_1, S_2, ..., S_n]$, if $1 \leq i < j \leq n$ (i.e., $S_i$ is a predecessor of $S_j$), then $State^{S_i}[k].seq \geq State^{S_j}[k].seq$.*

Formally, we restrict the history to one with a single value for each key and show that state transitions of NetChain can be mapped to those of the original CR, proving that NetChain provides per-key consistency. The proof shows that when a write query between switches is lost, it is equivalent to a state where the write query is processed by the tail switch, but the item is overwritten by a later write query, before the first query is exposed by any read queries. Moreover, as explained in §5, under various failure conditions, the protocol ensures that for each key, the versions exposed to client read queries are monotonically increasing. We provide a TLA+ verification of our protocol in the Appendix.

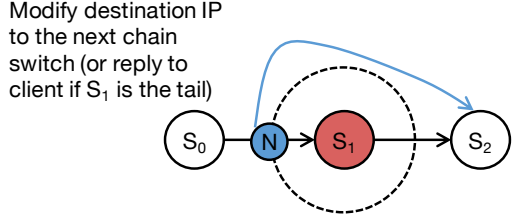

Figure 6: Fast failover.

## 5 NetChain Control Plane

**Overview.** The NetChain controller runs as a component in the network controller and only manages switch tables and registers related to NetChain. As discussed in §3, we assume the network controller is reliable. We mainly consider system reconfigurations caused by switch failures, which are detected by the network controller using existing techniques. We assume the failure model is fail-stop, and the switch failures can be correctly detected by the controller. To gracefully handle switch failures, we divide the process into two steps: *fast failover* and *failure recovery*. (*i*) In fast failover, the controller quickly reconfigures the network to resume serving queries with the remaining $f$ nodes in each affected chain. This degrades an affected chain to tolerate $f-1$ node failures. (*ii*) In failure recovery, the controller adds other switches as new replication nodes to the affected chains, which restores these chains to $f+1$ nodes. Since failure recovery needs to copy state to the new replicas, it takes longer than fast failover.

Other types of chain reconfigurations that (temporarily) remove switches from the network (e.g., switch firmware upgrade) are handled similarly to fast failover; those that add switches to the network (e.g., new switch onboarding) are handled similarly to failure recovery.

### 5.1 Fast Failover

Fast failover quickly removes failed switches and minimizes the durations of service disruptions caused by switch failures. Given a chain $[S_1, S_2, ..., S_k]$, when $S_i$ fails, the controller removes $S_i$ and the chain becomes $[S_1, ..., S_{i-1}, S_{i+1}..., S_k]$. A strawman approach is to notify the agents on all servers so that when the agents send out requests, they would put IPs of the updated chains to the packet headers. The drawback is the huge cost of disseminating chain updates to hundreds of thousands of servers in a datacenter. Another solution is to update $S_i$'s previous hop $S_{i-1}$. Specifically, after $S_{i-1}$ finishes processing the requests, it pops up one more IP address from the chain IP list and uses $S_{i+1}$ as the next hop. Given $n$ switches and $m$ virtual nodes, each switch is mapped to $m/n$ virtual nodes. Since each virtual node is in $f+1$ chains, a switch failure affects $m(f+1)/n$ chains in total. This implies that we need to update $m(f+1)/n$ switches

**Algorithm 2** Failover(fail_sw)

1:  **for** $sw \in fail\_sw.neighbors()$ **do**
2:      $rule.match \leftarrow dst\_ip = fail\_sw$
3:      **if** $fail\_sw$ is not tail **then**
4:          $rule.action \leftarrow$ (dst_ip = chain_ip[1], pop two chain IPs)
5:      **else**
6:          $rule.action \leftarrow$ (swap src_ip & dst_ip, pop chain IP)
7:      $sw.insert(rule)$

for one switch failure in fast failover. While $m(f+1)/n$ is fewer than the number of servers, it still incurs considerable reconfiguration cost as $m/n$ can be a few tens or hundreds. Furthermore, if $S_i$ is the head (tail) of a chain, the *previous* hop for write (read) queries would be *all the servers*, which need to be updated. In order to minimize service disruptions, we want to *reduce the number of nodes that must be updated* during failover.

**Reducing number of updated nodes.** Our key idea is that the controller only needs to update the *neighbor switches* of a failed switch to remove it from *all* its chains (Algorithm 2). Specifically, the controller inserts a new rule to each neighbor switch which examines the destination IP. If the destination IP is that of the failed switch (line 2), the neighbor switches copy the IP of the next chain hop of the failed switch (i.e., $S_{i+1}$'s IP) to the destination IP (line 3-4), or reply to the client if the failed switch is the tail (line 5-6). The condition of whether the failed switch is the tail (line 3) is implemented by checking the size of the chain IP list. Write and read queries are handled in the same way as they store the original and reverse orders of the chain IP list respectively. Figure 6 illustrates this idea. All requests from $S_0$ to $S_1$ have to go through $S_1$'s neighbor switches, which are represented by the dashed circle. $N$ is one of the neighbor switches. After $S_1$ fails, the controller updates $N$ to use $S_2$'s IP as the new destination IP. Note that if $N$ overlaps with $S_0$ ($S_2$), it updates the destination IP after (before) it processes the query. If $S_1$ is the end of a chain, $N$ would send the packet back to the client.

With this idea, even if a failed switch is the head or tail of a chain, the controller only needs to update its neighbor switches, instead of updating all the clients. Multiple switch failures are handled in a similar manner, i.e., updating the neighbor switches of the failed switches, and NetChain can only handle up to $f$ node failures for a chain of $f+1$ nodes.

### 5.2 Failure Recovery

Failure recovery restores all chains to $f+1$ switches. Suppose a failed switch $S_i$ is mapped to virtual nodes $V_1, V_2, ..., V_k$. These virtual nodes are removed from their chains in fast failover. To restore them, we first randomly assign them to $k$ live switches. This helps spread the load of failure recovery to multiple switches rather than a sin-gle switch. Let $V_x$ be reassigned to switch $S_y$. Since $V_x$ belongs to $f+1$ chains, we need to add $S_y$ to each of them. We use Figure 7 to illustrate how we add a switch to a chain. Fast failover updates $S_1$'s neighbor switches to forward all requests to $S_2$ (the blue line from $N$ to $S_2$). Failure recovery adds $S_3$ to the chain to restore the chain to three nodes (the orange lines from $N$ to $S_3$ and from $S_3$ to $S_2$). The process involves two steps, described below. Note that in essence the process is similar to live virtual machine migrations [28] and reconfigurations for server-based chain replication [24]. The difference is that we have the network controller to copy state and to modify switch rules to perform the reconfigurations. Algorithm 3 gives the pseudo code. In the example, $fail\_sw$, $new\_sw$ and $ref\_sw$ are $S_1$, $S_3$ and $S_2$, respectively.

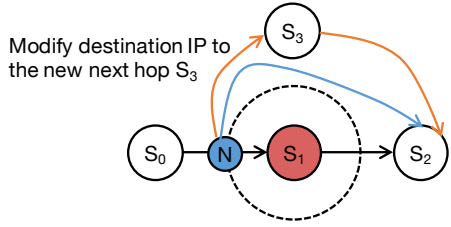

Figure 7: Failure recovery.

**Algorithm 3** FailureRecovery(fail_sw, new_sw)
1: $ref\_sw \leftarrow getLiveReferenceSwitch(fail\_sw)$
2: $preSyncState(new\_sw, ref\_sw)$
3: **for** $sw \in fail\_sw.neighbors()$ **do**
4: $\quad sw.stopForward(fail\_sw)$
5: $syncState(new\_sw, ref\_sw)$
6: $new\_sw.activateProcess()$
7: **for** $sw \in fail\_sw.neighbors()$ **do**
8: $\quad sw.activateForward(fail\_sw, new\_sw)$

**Step 1: Pre-synchronization.** The controller copies state from $S_2$ to $S_3$ (line 2). This includes (*i*) copying values from $S_2$'s register array to $S_3$'s register array, and (*ii*) inserting rules to $S_3$'s index table (Figure 3). While this step is time-consuming, the availability is not affected as the chain continues its operation with $S_0$ and $S_2$. We cover the case where there is no $S_2$ in the paragraph below on handling special cases.

**Step 2: Two-phase atomic switching.** The controller switches the chain from $[S_0, S_2]$ to $[S_0, S_3, S_2]$. This step has to be done carefully to ensure consistency. We keep the following invariant: *any node in the chain has newer values than its next hop.* The switching has two phases.

- **Phase 1: Stop and synchronization.** The controller inserts rules to *all* neighbors of $S_1$ ($N$ in Figure 7) to stop forwarding queries to $S_2$ (line 3-4). At the same time, it continues synchronizing the state between $S_2$ and $S_3$ (line 5). Since *no* new queries are forwarded to $S_2$, the state on $S_2$ and $S_3$ would eventually become the same. In this phase, the chain stops serving write queries, but still serves read queries with $S_2$.
- **Phase 2: Activation.** The controller activates the new chain switch $S_3$ to start processing queries (line 6), and activates *all* neighbor switches of $S_1$ ($N$ in the figure) to forward queries to $S_3$ (line 7-8). These rules modify the destination IP to $S_3$'s IP. They override the rules of fast failover (by using higher rule priorities), so that instead of being forwarded to $S_2$, all queries would be forwarded to $S_3$. The chain is restored after this phase.

Note that Step 1 is actually an optimization to shorten the temporary downtime caused by Phase 1 in Step 2, as most data is synchronized between the switches after Step 1. The invariant is kept throughout Phase 1 and Phase 2: in Phase 1, $S_3$ is synchronized to the same state as $S_2$ before it is added to the chain; in Phase 2, the neighbor switches of $S_1$ gradually restart to forward write queries again, which always go to $S_3$ first. The read queries in this process are continuously handled by $S_2$ and do not have temporary disruptions as write queries. The failure recovery process is performed for each failed switch under multiple failures.

**Handling special cases.** The above example shows the normal case when $S_1$ is in the middle of a chain. Now we discuss the special cases when $S_1$ is the head or tail of a chain. (*i*) When $S_1$ is the head, the process is the same as $S_1$ is in the middle. In addition, since the head switch assigns sequence numbers, the new head must assign sequence numbers bigger than those assigned by the failed head. To do this, we adopt a mechanism similar to NOPaxos [20], which uses an additional session number for message ordering. The session number is increased for every new head of a chain, and the messages are ordered by the lexicographical order of the (session number, sequence number) tuple. (*ii*) When $S_1$ is the tail, in Step 1, we copy state from its previous hop to it as we do not have $S_2$ (line 1 would assign $S_0$ to $ref\_sw$). In Phase 1 of Step 2, both read and write queries are dropped by $S_1$'s neighbors to ensure consistency. Although $S_0$ is still processing write queries during synchronization, we only need to copy the state from $S_0$ that are updated before we finish dropping all queries on $S_1$'s neighbor switches, as no new read queries are served. Then in Phase 2, we activate the chain by forwarding the queries to $S_3$.

**Minimizing disruptions with virtual groups.** While we guarantee the consistency for failure recovery with two-phase atomic switching, write queries need to be stopped to recover head and middle nodes, and both read and write queries need to be stopped to recover tail nodes. We use virtual groups to minimize the service disruptions. Specifically, we only recover one virtual group each time. Let a switch be mapped to 100 virtual groups. Each group is available by 99% of the recovery time and only queries to one group are affected at each time.

## 6 Discussion

**Switch on-chip storage size.** Coordination services are not intended to provide generic storage services. We argue that the switch on-chip memory is sufficient from two aspects. (*i*) *How much can the network provide?* Commodity switches today have tens of megabytes of on-chip SRAM. Because datacenter networks do not use many protocols as WANs and use shallow buffers for low latency, a large portion of the on-chip SRAM can be allocated to NetChain. Assuming 10MB is allocated in each switch, a datacenter with 100 switches can provide 1GB total storage, or 333MB effective storage with a replication factor of three. The number can be further increased with better semiconductor technology and allocating more on-chip SRAM to NetChain. (*ii*) *How much does NetChain need?* As shown in [1], a Chubby instance at Google that serves tens of thousands of clients store 22k files in total, among which ∼90% are 0-1 KB and only ∼0.2% are bigger than 10KB. This means 42MB is enough to store ∼99.8% of the files. Furthermore, consider the use case for locks. Assuming each lock requires 30B, then 333MB storage can provide 10 million concurrent locks. For in-memory distributed transactions that take $100\mu s$, NetChain would be able to provide 100 billion locks per second, which is enough for the most demanding systems today. Even with a small deployment using three switches (providing 10MB storage), NetChain would be able to provide 0.3 million concurrent locks or 3 billion locks per second.

**Value size.** The value size is limited by the packet size (9KB for Ethernet jumbo frames). Conceivably, a big value can be stored with multiple keys, but strong consistency is not provided for a multi-key query spanning several packets. The value size is also limited by the switch chip. Typically, a switch pipeline contains multiple stages ($k$) and each stage can read or write a few bytes ($n$). Assuming $k$=12 and $n$=16, switches can handle values up to $kn$=192 bytes at line rate. Values bigger than $kn$ can be supported using packet mirroring/recirculation which sends packets to go through the pipeline for another round of processing, but at the cost of lower effective throughput [18]. We suggest that NetChain is best suitable for small values that need frequent access, such as configuration parameters, barriers and locks.

**Data placement.** NetChain uses consistent hashing and virtual nodes to partition the key-value store between multiple switches. The data placement strategy can be optimized for throughput and latency by taking into account the network topology and the query characteristics.

**Full interface.** The current NetChain prototype provides a basic key-value interface for fixed-length keys and limited-size variable-length values. Many commercial and open-source systems like ZooKeeper provide additional features, e.g., hierarchical name space (as a file system), watches (which notify clients when watched values are updated), access control list (ACL) and data encryption. We leave these features as future work.

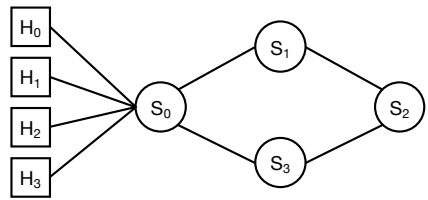

Figure 8: NetChain testbed with four 6.5Tbps Barefoot Tofino switches ($S_0$-$S_3$) and four servers ($H_0$-$H_3$).

**Accelerator for server-based solutions.** NetChain can be used as an accelerator to server-based solutions such as Chubby [1], ZooKeeper [2] and etcd [3]). The key space is partitioned to store data in the network and the servers separately. NetChain can be used to store *hot* data with *small* value size, and servers store *big* and *less popular* data. Such a hybrid approach provides the advantages of both solutions.

## 7 Implementation

We have implemented a prototype of NetChain, including a switch data plane, a controller and a client agent. (*i*) The switch data plane is written in P4 [29] and is compiled to Barefoot Tofino ASIC [13] with Barefoot Capilano software suite [30]. We use 16-byte keys and use 8 stages to store values. We allocate 64K 16-byte slots in each stage. This in total provides 8 MB storage. We use standard L3 routing that forwards packets based on destination IP. In total, the NetChain data plane implementation uses much less than 50% of the on-chip memory in the Tofino ASIC. (*ii*) The controller is written in Python. It runs as a process on a server and communicates with each switch agent through the standard Python RPC library xmlrpclib. Each switch agent is a Python process running in the switch OS. It uses a Thrift API generated by the P4 compiler to manage the switch resources and update the key-value items in the data plane (e.g., for failure handling) through the switch driver. (*iii*) The client agent is implemented in C with Intel DPDK [31] for optimized IO performance. It provides a key-value interface for applications, and achieves up to 20.5 MQPS with the 40G NICs on our servers.

## 8 Evaluation

In this section, we provide evaluation results to demonstrate NetChain provides orders of magnitude improvements on throughput (§8.1) and latency (§8.2), is scalable (§8.3), handles failures gracefully (§8.4), and significantly benefits applications (§8.5).

**Testbed.** Our evaluation is conducted on a testbed consisting of four 6.5 Tbps Barefoot Tofino switches and four server machines. Each server machine is equipped with a 16-core CPU (Intel Xeon E5-2630) and 128 GB total memory (four Samsung 32GB DDR4-2133 memory). Three server machines are equipped with 40G NICs (Intel XL710) and the other one is equipped with a 25G NIC (Intel XXV710). The testbed is organized in a topology as shown in Figure 8.

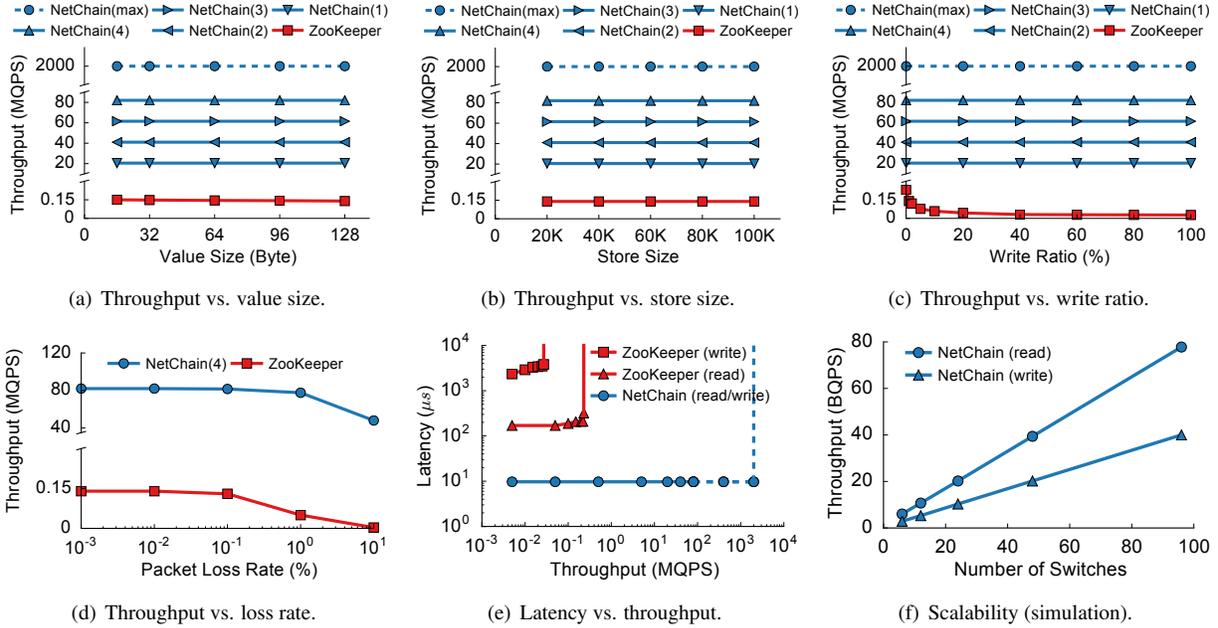

Figure 9: Performance results. (a-e) shows the experimental results of a three-switch NetChain prototype. Netchain(1), Netchain(2), Netchain(3) and Netchain(4) correspond to measuring the prototype performance with one, two, three and four servers respectively. NetChain(max) is the theoretical maximum throughput achievable by a three-switch chain; it is not a measured throughput. (f) shows the simulation results of spine-leaf networks of various sizes.

**Comparison.** We compare NetChain to Apache ZooKeeper-3.5.2 [32]. We implement a client to measure ZooKeeper's performance with Apache Curator-4.0.0 [33], which is a popular client library for ZooKeeper. The comparison is slightly unfair: NetChain does not provide all features of ZooKeeper (§6), and ZooKeeper is a production-quality system that compromises its performance for many software-engineering objectives. But at a high level, the comparison uses ZooKeeper as a reference for server-based solutions to demonstrate the performance advantages of NetChain.

## 8.1 Throughput

We first evaluate the throughput of NetChain. We use three switches to form a chain $[S_0, S_1, S_2]$, where $S_0$ is the head and $S_2$ is the tail. Each server can send and receive queries at up to 20.5 MQPS. We use NetChain(1), NetChain(2), NetChain(3), NetChain(4) to denote the measured throughput by using one, two, three and four servers, respectively. We use Tofino switches in a mode that guarantees up to 4 BQPS throughput and each query packet is processed twice by a switch (e.g., a query from $H_0$ follows path $H_0$-$S_0$-$S_1$-$S_2$-$S_1$-$S_0$-$H_0$). Therefore, the maximum throughput of the chain is 2 BQPS in this setup. As the four servers cannot saturate the chain, we use NetChain(max) to denote the maximum throughput of the chain (shown as dotted lines in figures). For comparison, we run ZooKeeper on three servers, and a separate 100 client processes on the other server to generate queries. This experiment aims to thoroughly evaluate the throughput of one switch chain under various setups with real hardware switches. For large-scale deployments, a packet may traverse multiple hops to get from one chain switch to the next, and we evaluate the throughput with simulations in §8.3. Figure 9(a-d) shows the throughputs of the two systems. The default setting uses 64-byte value size, 20K store size (i.e., the number of key-value items), 1% write ratio, and 0% link loss rate. We change one parameter in each experiment to show how the throughputs are affected by these parameters.

Figure 9(a) shows the impact of value size. NetChain provides orders of magnitude higher throughput than ZooKeeper and both systems are not affected by the value size in the evaluated range. NetChain(4) keeps at 82 MQPS, meaning that NetChain can fully serve all the queries generated by the four servers. This is due to the nature of a switch ASIC: as long as the P4 program is compiled to fit the switch resource requirements, the switch is able to run NetChain at line rate. In fact, a three-switch chain is able to provide up to 2 BQPS, as denoted by NetChain(max). Our current prototype support value size up to 128 bytes. Larger values can be supported using more stages and using packet mirroring/recirculation as discussed in §6.

Figure 9(b) shows the impact of store size. Similarly, both systems are not affected by the store size in the evaluated range, and NetChain provides orders of magnitude higher throughput. The store size is restricted by the allocated total size (8MB in our prototype) and the value size. The store size is large enough to be useful for coordination services as discussed in §6.

Figure 9(c) shows the impact of write ratio. With read-only workloads, ZooKeeper achieves 230 KQPS. But even with a write ratio of 1%, its throughput drops to 140 KQPS. And when the write ratio is 100%, its throughput drops to 27 KQPS. As for comparison, NetChain(4) consistently achieves 82 MQPS. This is because NetChain uses chain replication and each switch is able to process both read and write queries at line rate. As the switches in the evaluated chain $[S_0, S_1, S_2]$ process the same number of packets for both read and write queries, the total throughput is not affected by the write ratio, which would be different in more complex topologies. As we will show in §8.3, NetChain has lower throughput for write queries for large deployments, as write queries require more hops than read queries.

Figure 9(d) shows the impact of packet loss rate. We inject random packet loss rate to each switch, ranging from 0.001% to 10%. The throughput of ZooKeeper drops to 50 KQPS (3 KQPS) when the loss rate is 1% (10%). As for comparison, NetChain(4) keeps around 82 MQPS for packet loss rate between 0.001% and 1%, and only drops to 48 MPQS when the loss rate is 10%. The reason is because ZooKeeper uses TCP for reliable transmission which has a lot of overhead under high loss rate, whereas NetChain simply uses UDP and lets the clients retry a query upon packet loss. Although high packet loss rate is unlikely to happen frequently in datacenters, this experiment demonstrates that NetChain can provide high throughput even under extreme scenarios.

### 8.2 Latency

We now evaluate the latency of NetChain. We separate the read and write queries, and measure their latencies under different throughputs. For NetChain, since the switch-side processing delay is sub-microsecond, the client-side delay dominates the query latency. In addition, as both read and write queries traverse the same number of switches in the evaluated chain, NetChain has the same latency for both reads and writes, as shown in Figure 9(e). Because we implement NetChain clients with DPDK to bypass the TCP/IP stack and the OS kernel, NetChain incurs only 9.7 $\mu$s query latency. The la-tency keeps at 9.7 $\mu$s even when all four severs are generating queries to the system at 82 MQPS (the solid line of NetChain in the figure), and is expected to be not affected by throughput until the system is saturated at 2 BQPS (the dotted line of NetChain in the figure).

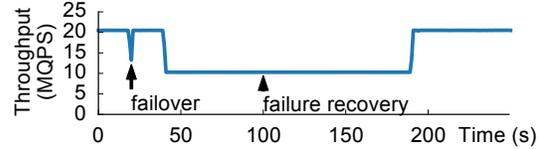

(a) 1 Virtual Group.

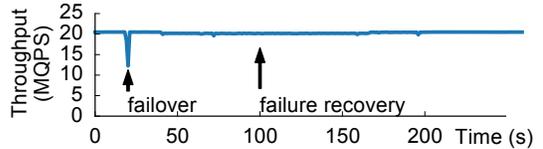

(b) 100 Virtual Groups.

Figure 10: Failure handling results. It shows the throughput time series of one client server when one switch fails in a four-switch testbed. NetChain has fast failover. By using more virtual groups, NetChain provides smaller throughput drops for failure recovery.

As for comparison, ZooKeeper has a latency of 170 $\mu$s for read queries and 2350 $\mu$s for write queries at low throughput. The latencies slightly go up before the system is saturated (27 KQPS for writes and 230 KQPS for reads), because servers do not have deterministic per-query processing time as switch ASICs and the latency is affected by the system load. Overall, NetChain provides orders of magnitude lower latency than ZooKeeper at orders of magnitude higher throughput.

### 8.3 Scalability

We use simulations to evaluate the performance of NetChain in large-scale deployments. We use standard spine-leaf topologies. We assume each switch has 64 ports and has a throughput of 4 BQPS. Each leaf switch is connected to 32 servers in its rack, and uses the other 32 ports to connect to spine switches. We assume the network is non-blocking, i.e., the number of spine switches is a half of that of leaf switches. We vary the network from 6 switches (2 spines and 4 leafs) to 96 switches (32 spines and 64 leafs). Figure 9(f) shows the maximum throughputs for read-only and write-only workloads. Both throughputs grow linearly, because in the two-layer network, the average number of hops for a query does not change under different network sizes. The write throughput is lower than the read throughput because a write query traverses more hops than a read query. When the queries have mixed read and write operations, the throughput curve will be between NetChain(read) and NetChain(write).

## 8.4 Handling Failures

The four-switch testbed allows us to evaluate NetChain under different failure conditions. For example, to evaluate the failure of a middle (tail) node, we can use an initial chain $[S_0,S_1,S_2]$, fail $S_1$ ($S_2$), and let $S_3$ replace $S_1$ ($S_2$). Since failing $S_0$ would disconnect the servers from the network, to evaluate a head failure and show the system throughput changes during the transition, we can use an initial chain $[S_1,S_2,S_0]$, fail $S_1$, and let $S_3$ replace $S_1$. Due to the limited space, we show one such experiment where we fail $S_1$ in the chain $[S_0,S_1,S_2]$ and use $S_3$ to replace $S_1$ for failure recovery. We use a write ratio of 50%, and let the write queries use path $S_0$-$S_1$-$S_2$ and the read queries use path $S_0$-$S_3$-$S_2$. In this way, we can demonstrate that when the middle node fails, the system is still able to serve read queries. We show the throughput time series of one client with different numbers of virtual groups.

Figure 10(a) shows the throughput time series with only one virtual group. Initially, the client has a throughput of 20.5 MQPS. Then at 20s, we inject the failure of $S_1$ by letting $S_0$ drop all packets to $S_1$. NetChain quickly fails over to a two-switch chain $[S_0,S_2]$, and the client throughput is fully restored. Note that to make the throughput drop visible, we manually inject a one-second delay to the controller before it starts the failover routine. In practice, the duration of failover depends on how fast the control plane can detect the failure, which is beyond the scope of this paper. We also add 20 seconds to separate the failure recovery process from the failover process. The failure recovery starts around 40s and lasts about 150s. The throughput drops to half because NetChain needs to synchronize the key-value store between $S_2$ and $S_3$, during which write queries cannot be served. Note that if it was a tail failure, the throughput would drop to 0, because both read and write queries cannot be served during the recovery of the tail (§5).

Figure 10(b) shows the time series of NetChain throughput with 100 virtual groups. The failover part is similar, but the failure recovery part only experiences a 0.5% throughput drop. This is because by using 100 virtual groups, NetChain only needs to stop serving write queries for one virtual group each time, or 1% of queries. Since we use a write ratio of 50%, only $1\% \times 50\% = 0.5\%$ of queries are affect during failure recovery. Therefore, using more virtual groups provides both slight throughput drops and better availability.

## 8.5 Application Performance

Finally, we use distributed transactions as an application to further demonstrate the benefits of NetChain. We use a benchmark workload from previous work [34, 35], which is a generalization of the new-order transaction in TPC-C benchmark [36], because the workload allows us to test transactions with different contention levels. In this workload, for each transaction, a client needs to acquire ten locks, where one lock is chosen from a small set of hot items and the other nine locks are from a very large set of items. The workload uses *contention index*, which is the inverse of the number of hot items, to control the contentions. For example, a contention index of 0.001 (or 1) means all clients compete for one item in 1000 (or 1) items. We use the classic two-phase locking (2PL) protocol: each client first acquire all the locks from NetChain or ZooKeeper, and then releases all the locks to complete one transaction. For NetChain, we use the compare-and-swap (CAS) primitive in Tofino switches to implement exclusive locks. Specifically, in addition to checking sequence numbers, a lock can only be released by the client that owns the lock by comparing the client ID in the value field. For ZooKeeper, exclusive locks can be implemented by ephemeral znodes and are directly provided by Apache Curator client library.

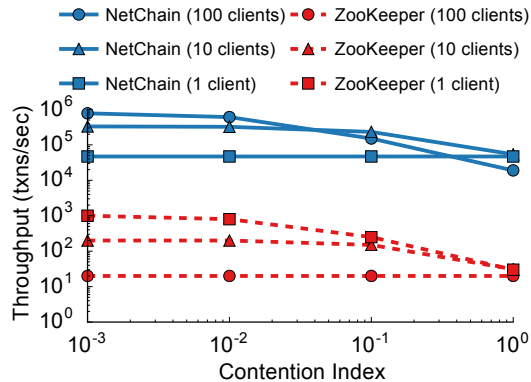

Figure 11: Application results. NetChain provides orders of magnitude higher transaction throughput.

Figure 11 shows the transaction throughputs. By using NetChain as a locking server, the system can achieve orders of magnitude higher transaction throughput than ZooKeeper. We observe that the line with one client is flat because there are no contentions with one client. With 100 clients, the system has higher throughput at small contention index, because more clients can do more transactions. But the throughput decreases when the contention index increases, and is even slightly lower than that with one client, due to more contentions. We expect a lightning fast coordination system like NetChain can open the door for designing a new generation of distributed systems beyond distributed transactions.

## 9 Related Work

**Consensus protocols.** Various protocols have been proposed to solve the distributed consensus problem, such as Paxos [4], ZAB [37], Raft [38], Viewstamped Replication [39], Virtual Synchrony [40], etc. By leveraging that in some scenarios messages arrive in order, some protocols are designed to reduce the overhead, such as Fast Paxos [41] and Optimistic Atomic Broadcast [42]. Recent work goes one step further by ordering messages with the network, including SpecPaxos [43] and NOPaxos [20]. NetChain directly processes coordination queries in the network, providing higher performance.

**Coordination services.** As distributed coordination is widely used, some systems are built to provide coordination as a service, e.g., Chubby [1], ZooKeeper [2] and etcd [3]. These services have highly-optimized implementations of consensus protocols in their core and provide simple APIs. NetChain provides a similar key-value API and implements the system in the network.

**Hardware accelerations.** NetPaxos [16, 17] implements Paxos on switches. It still requires to build a replicated key-value store on servers, which may use NetPaxos for consensus to improve performance. As such, the key-value store is still bounded by server IO, and thus is much slower than NetChain. Besides, NetChain uses Vertical Paxos for consensus which is more suitable for in-network key-value stores, provides protocols and algorithms for routing and failure handling, and has an implementation with multiple switches and an evaluation. Compared to NetCache [18], NetChain uses NetCache's on-chip key-value store design, and designs a replicated, in-network key-value store that handles both read and write queries and provides strong consistency and fault-tolerance. Some other work has also used hardware to speed up distributed systems. SwitchKV [44] uses switches to enable content-based routing for load balancing of key-value stores; the switches do not cache key-value items or serve queries. Marple [45] designs a new hardware primitive to support key-value stores for network measurements. Schiff *et al.* [46] designs a synchronization framework to resolve conflicts for distributed network controllers, which has to go through the switch control plane. Li *et al.* [47] proposes a new hardware design to achieve a throughput of 1 BQPS with a single server platform. IncBricks [48] caches key-value items with NPUs. Consensus in a box [49] uses FPGAs to speed up ZooKeeper's atomic broadcast protocol. NetChain does not require specialized chips, and switch ASICs have higher performance than NPUs and FPGAs.

## 10 Conclusion

We present NetChain, an in-network coordination service that provides billions of coordination operations per second with sub-RTT latencies. NetChain leverages programmable switches to build a strongly-consistent, fault-tolerant, in-network key-value store. This powerful capability dramatically reduces the coordination latency to as little as half of an RTT. We believe NetChain exemplifies a new generation of ultra-low latency systems enabled by programmable networks.

**Acknowledgments** We thank our shepherd Amar Phanishayee and the anonymous reviewers for their valuable feedback. Robert Soulé is supported in part by SNF 167173. Nate Foster is supported in part by NSF CNS-1413972. Ion Stoica is supported in part by DHS Award HSHQDC-16-3-00083, NSF CISE Expeditions Award CCF-1139158, and gifts from Ant Financial, Amazon Web Services, CapitalOne, Ericsson, GE, Google, Huawei, Intel, IBM, Microsoft and VMware.

# Appendix. TLA+ Verification

---- MODULE *NetChain* ----

Specifies the *NetChain* request handling process

EXTENDS *Naturals*, *Sequences*, *FiniteSets*

Assuming finite number of requests and buffer length for model checking
CONSTANTS Requests,                \* List of requests, assigned in model
CONSTANTS $maxQLen$,               Maximum length of channel buffer
          $maxFailedCount$,        Maximum number of failed switches
          $maxVersion$,            Maximum version number
          $maxBufOpCount$          Maximum number of buffer operations

CONSTANTS *Switches*,              Set of *NetChain* switches
          *Keys*,                  Key space
          *Values*,                Value space
          *SetNodes*,              All nodes including switches and clients
          *SetValues*              All possible values

Switch state.
VARIABLES $sMem$,                  Switch internal memory
          $sBuf$,                  Switch incoming message buffer
          $sBufSt$,                Switch incoming buffer status
          $sStatus$,               Switch liveness status
          $sReadFwd$,              Switch forward hop for read if failed
          $sWriteFwd$,             Switch forward hop for write if failed
          $sFailedCount$           Number of failed switches

Communication channel state. Each channel contains a single direction
buffer which holds the messages from *src* to *dst*.
VARIABLES $allBuf$,                Channel message buffer
          $allBufOp$,              Indicates whether operation was performed
          $bufOpCount$             Total number of buffer operations

Client observed key-values.
VARIABLES $prevKVs$,                Last observed key-values
          $currentKVs$              Current observed key-values

---

Helper functions.

$BoundedSeq(Data, N) \triangleq \text{UNION } \{[1 .. n \to Data] : n \in 0 .. N\}$

$Range(T) \triangleq \{T[x] : x \in \text{DOMAIN } T\}$

$IsElementInSeq(el, seq) \triangleq \exists i \in \text{DOMAIN } seq : seq[i] = el$

Get a value that is not used in any requests to initialize memory.
$NoVal \triangleq \text{CHOOSE } v \in SetValues : v \notin Values$

Get a node that is not a switch for client. This client can have a lot of outstanding requests, simulating a lot of client nodes.
$Client \triangleq \text{CHOOSE } h \in SetNodes : h \notin Switches$

All active nodes in the system.
$AllNodes \triangleq Switches \cup \{Client\}$

All values including $NoVal$.
$AllValues \triangleq Values \cup \{NoVal\}$

Define $NetChain$: a chain of distinct switches.
$SwitchChain \triangleq \text{CHOOSE } chain \in BoundedSeq(Switches, Cardinality(Switches)) :$
$\qquad\qquad\qquad Len(chain) = Cardinality(Range(chain))$
$\qquad\qquad\qquad \wedge\ Len(chain) = Cardinality(Switches) - maxFailedCount$

$InnerSwitchToIdx[s \in Range(SwitchChain),$
$\qquad\qquad chain \in BoundedSeq(Switches, Len(SwitchChain))] \triangleq$
$\quad \text{IF } s = Head(chain) \text{ THEN } 1 \text{ ELSE } 1 + InnerSwitchToIdx[s, Tail(chain)]$

Return the index of a switch in $SwitchChain$
$SwitchToIdx(s) \triangleq \text{IF } \neg IsElementInSeq(s, SwitchChain)$
$\qquad\qquad\qquad \text{THEN } 0$
$\qquad\qquad\qquad \text{ELSE } InnerSwitchToIdx[s, SwitchChain]$

Get the next switch in chain. If the given switch is not in chain or it's the tail switch, return $x$.
$NextInSeqOrX(s, x) \triangleq \text{IF } \neg IsElementInSeq(s, SwitchChain)$
$\qquad\qquad\qquad\qquad \vee\ s = SwitchChain[Len(SwitchChain)]$
$\qquad\qquad\qquad \text{THEN } x$
$\qquad\qquad\qquad \text{ELSE } SwitchChain[SwitchToIdx(s) + 1]$

Get the previous switch in chain. If the given switch is not in chain or it's the head switch, return $x$.
$PrevInSeqOrX(s, x) \triangleq \text{IF } \neg IsElementInSeq(s, SwitchChain)$
$\qquad\qquad\qquad\qquad \vee\ s = SwitchChain[1]$
$\qquad\qquad\qquad \text{THEN } x$
$\qquad\qquad\qquad \text{ELSE } SwitchChain[SwitchToIdx(s) - 1]$

Get the next alive switch in chain, or return $Client$ if no such switch.
$NextAliveInSeqOrClient[s \in Range(SwitchChain)] \triangleq$
$\quad \text{LET } next \triangleq NextInSeqOrX(s, Client)$
$\quad \text{IN } \text{ IF } next = Client$
$\qquad\qquad \text{THEN } Client$

     ELSE IF $sStatus[next] =$ "alive"
     THEN $next$
     ELSE IF $sStatus[next] =$ "recovered"
     THEN $sWriteFwd[next]$
     ELSE $NextAliveInSeqOrClient[next]$

Get the previous alive switch in chain, or return *Client* if no such switch.
$PrevAliveInSeqOrClient[s \in Range(SwitchChain)] \triangleq$
 LET $prev \triangleq PrevInSeqOrX(s, Client)$
 IN IF $prev = Client$
    THEN $Client$
    ELSE IF $sStatus[prev] =$ "alive"
    THEN $prev$
    ELSE IF $sStatus[prev] =$ "recovered"
    THEN $sWriteFwd[prev]$
    ELSE $PrevAliveInSeqOrClient[prev]$

4 message types for *Read* requests, *Write* requests, *Read/Write* replies,
and *NoOp* messages. The *val* field is a (value, version) tuple.
$ReadReq \triangleq [op : \{\text{"Read"}\}, key : Keys, hops : Seq(Switches)]$
$WriteReq \triangleq [op : \{\text{"Write"}\}, key : Keys, val : AllValues \times Nat, hops : Seq(Switches)]$
$ReplyMsg \triangleq [op : \{\text{"Reply"}\}, key : Keys, val : AllValues \times Nat]$
$NoOpMsg \triangleq [op : \{\text{"NoOp"}\}]$
$Messages \triangleq ReadReq \cup WriteReq \cup ReplyMsg \cup NoOpMsg$

---

Initial state.

All switch memory with *NoVal*;

All empty incoming buffers ("rdy" indicates buffer is empty);

Status of all switches to be alive;

Forward nodes set to *Client* (indicating invalid forward switch)
$InitSMem \triangleq \; \wedge sMem = [s \in Switches \mapsto [k \in Keys \mapsto \langle NoVal, 0 \rangle]]$
      $\wedge sBuf \; = [s \in Switches \mapsto [op \mapsto \text{"NoOp"}]]$
      $\wedge sBufSt = [s \in Switches \mapsto \text{"rdy"}]$
      $\wedge sStatus = [s \in Switches \mapsto \text{"alive"}]$
      $\wedge sReadFwd = [s \in Switches \mapsto Client]$
      $\wedge sWriteFwd = [s \in Switches \mapsto Client]$
      $\wedge sFailedCount = 0$

All communication channels between nodes are empty;

All channels have no operations (drop, repeat, reorder) performed
$InitBufs \triangleq \; \wedge allBuf = [n \in AllNodes \mapsto [m \in AllNodes \mapsto \langle \rangle]]$
      $\wedge allBufOp = [n \in AllNodes \mapsto [m \in AllNodes \mapsto \text{FALSE}]]$
      $\wedge bufOpCount = 0$

External (client) observed memory.

$InitCMem \triangleq \wedge prevKVs = [k \in Keys \mapsto \langle NoVal, 0\rangle]$
$\qquad\qquad\quad \wedge currentKVs = [k \in Keys \mapsto \langle NoVal, 0\rangle]$

$Init \triangleq \wedge InitSMem$
$\qquad\quad \wedge InitBufs$
$\qquad\quad \wedge InitCMem$

---

Type Invariant.

$SMem\_TypeInvariant \triangleq$
$\quad \wedge sMem \in [Switches \to [Keys \to AllValues \times Nat]]$
$\quad \wedge sBuf \;\in [Switches \to Messages]$
$\quad \wedge sBufSt \in [Switches \to \{\text{"rdy"}, \text{"busy"}\}]$
$\quad \wedge sStatus \in [Switches \to \{\text{"alive"}, \text{"failed"}, \text{"recovered"}\}]$
$\quad \wedge sReadFwd \in [Switches \to AllNodes]$
$\quad \wedge sWriteFwd \in [Switches \to AllNodes]$

$Buf\_TypeInvariant \triangleq$
$\quad \wedge allBuf \in [AllNodes \to [AllNodes \to Seq(Messages)]]$
$\quad \wedge allBufOp \in [AllNodes \to [AllNodes \to \text{BOOLEAN}]]$

$CMem\_TypeInvariant \triangleq$
$\quad \wedge prevKVs \in [Keys \to AllValues \times Nat]$
$\quad \wedge currentKVs \in [Keys \to AllValues \times Nat]$

$TypeInvariant \triangleq \wedge SMem\_TypeInvariant$
$\qquad\qquad\qquad\quad \wedge Buf\_TypeInvariant$
$\qquad\qquad\qquad\quad \wedge CMem\_TypeInvariant$

---

Channel related operators

Send a message: update buffer and buffer operation status

$BufSend(src, dst, msg) \triangleq$
$\quad \wedge \; allBuf' = [allBuf \text{ EXCEPT } ![src][dst] = Append(allBuf[src][dst], msg)]$
$\quad \wedge \; allBufOp' = [allBufOp \text{ EXCEPT } ![src][dst] = \text{FALSE}]$

Receive a message: remove the message from queue, and reset the operation flag. Caller should copy the message from head of queue to buffer.

$BufRcv(src, dst) \triangleq$
$\quad \wedge allBuf[src][dst] \neq \langle\rangle$
$\quad \wedge allBuf' = [allBuf \text{ EXCEPT } ![src][dst] = Tail(allBuf[src][dst])]$
$\quad \wedge allBufOp' = [allBufOp \text{ EXCEPT } ![src][dst] = \text{FALSE}]$

Drop, repeat, or reorder messages in buffer.

$BufDrop(src, dst) \triangleq$
    $allBuf' = [allBuf \text{ EXCEPT } ![src][dst] = Tail(allBuf[src][dst])]$
$BufRepeat(src, dst) \triangleq$
    $allBuf' = [allBuf \text{ EXCEPT}$
        $![src][dst] = Append(allBuf[src][dst], Head(allBuf[src][dst]))]$
$BufReorder(src, dst) \triangleq$
    $allBuf' = [allBuf \text{ EXCEPT}$
        $![src][dst] = Append(Tail(allBuf[src][dst]), Head(allBuf[src][dst]))]$

Buffer performs an operation (drop, repeat, reorder)
$BufOp(src, dst) \triangleq$
    $\wedge\ allBuf[src][dst] \neq \langle\rangle$     Enable only if q is nonempty
    $\wedge\ allBufOp[src][dst] = \text{FALSE}$     and no operation was performed.
    $\wedge\ (BufDrop(src, dst) \vee BufRepeat(src, dst) \vee BufReorder(src, dst))$
    $\wedge\ allBufOp' = [allBufOp \text{ EXCEPT } ![src][dst] = \text{FALSE}]$
    $\wedge\ bufOpCount' = bufOpCount + 1$
    $\wedge\ \text{UNCHANGED}\ \langle sMem, sBuf, sBufSt, sStatus, sReadFwd, sWriteFwd,$
                      $sFailedCount, prevKVs, currentKVs\rangle$

---

Message processing in $NetChain$ nodes.

Switch receives a message from upstream channels (sent by other nodes) and
put it inside the incoming buffer
$SwitchRcv(s) \triangleq$
    $\wedge\ sBufSt[s] = \text{"rdy"}$
    $\wedge\ \exists n \in AllNodes:$
        $(\wedge\ BufRcv(n, s)$
          $\wedge\ sBuf' = [sBuf \text{ EXCEPT } ![s] = Head(allBuf[n][s])]$
          $\wedge\ sBufSt' = [sBufSt \text{ EXCEPT } ![s] = \text{"busy"}])$
    $\wedge\ \text{UNCHANGED}\ \langle sMem, sStatus, sReadFwd, sWriteFwd, sFailedCount,$
                      $prevKVs, currentKVs, bufOpCount\rangle$

For read request, directly reply to client
$SProcRead(s) \triangleq$
    $\wedge\ sBuf[s].op = \text{"Read"}$
    $\wedge\ BufSend(s, Client, [op \mapsto \text{"Reply"},$
                       $key \mapsto sBuf[s].key,$
                       $val \mapsto sMem[s][sBuf[s].key]])$
    $\wedge\ \text{UNCHANGED}\ sMem$

For write request, get the message, perform the write (if seeing newer
versions), and forward it to next hop or reply to client
$SProcWrite(s) \triangleq$
    $\wedge\ sBuf[s].op = \text{"Write"}$
    $\wedge\ \text{LET}\ m\ \triangleq\ sBuf[s]$

$$
\begin{aligned}
&ver \overset{\Delta}{=} \text{IF } m.val[2] = 0 \\
&\qquad\qquad \text{THEN } sMem[s][m.key][2] + 1 \\
&\qquad\qquad \text{ELSE } m.val[2] \\
&\text{IN} \quad \text{IF } ver > sMem[s][m.key][2] \qquad \text{process only newer versions}\\
&\qquad \text{THEN } (sMem' = [ss \in Switches \mapsto \\
&\qquad\qquad\qquad\qquad \text{IF } ss = s \\
&\qquad\qquad\qquad\qquad\qquad \text{THEN } [sMem[ss] \text{ EXCEPT } ![m.key] = \langle m.val[1], ver\rangle] \\
&\qquad\qquad\qquad\qquad\qquad \text{ELSE } sMem[ss]] \\
&\qquad\qquad \wedge \text{IF } Len(m.hops) > 1 \\
&\qquad\qquad\qquad \text{THEN } BufSend(s, m.hops[2], [op \mapsto m.op, \\
&\qquad\qquad\qquad\qquad\qquad\qquad\qquad\qquad key \mapsto m.key, \\
&\qquad\qquad\qquad\qquad\qquad\qquad\qquad\qquad val \mapsto \langle m.val[1], ver\rangle, \\
&\qquad\qquad\qquad\qquad\qquad\qquad\qquad\qquad hops \mapsto Tail(m.hops)]) \\
&\qquad\qquad\qquad \text{ELSE } BufSend(s, Client, [op \mapsto \text{``Reply''}, \\
&\qquad\qquad\qquad\qquad\qquad\qquad\qquad\qquad key \mapsto m.key, \\
&\qquad\qquad\qquad\qquad\qquad\qquad\qquad\qquad val \mapsto \langle m.val[1], ver\rangle])) \\
&\qquad \text{ELSE UNCHANGED } \langle sMem, allBuf, allBufOp \rangle
\end{aligned}
$$

For a switch that is alive, use above read / write logic

$SwitchAliveProc(s) \overset{\Delta}{=}$
$\quad \wedge sBufSt[s] = \text{``busy''} \wedge sStatus[s] = \text{``alive''}$
$\quad \wedge (SProcRead(s) \vee SProcWrite(s))$
$\quad \wedge sBufSt' = [sBufSt \text{ EXCEPT } ![s] = \text{``rdy''}]$
$\quad \wedge \text{UNCHANGED } \langle sBuf, sStatus, sReadFwd, sWriteFwd, sFailedCount,$
$\qquad\qquad\qquad\qquad prevKVs, currentKVs, bufOpCount \rangle$

$SFailedRead(s) \overset{\Delta}{=}$
$\quad \wedge sBuf[s].op = \text{``Read''}$
$\quad \wedge BufSend(s, sReadFwd[s], sBuf[s])$

Write request on failed switch:

case 1: not tail switch, or switch is already recovered

case 2: failed tail switch and not yet recovered

$SFailedWrite(s) \overset{\Delta}{=}$
$\quad \wedge sBuf[s].op = \text{``Write''}$
$\quad \wedge \text{LET } m \overset{\Delta}{=} sBuf[s]$
$\qquad \text{IN} \quad \text{IF } sStatus[s] = \text{``recovered''}$
$\qquad\qquad \text{THEN } BufSend(s, sWriteFwd[s], m)$
$\qquad\qquad \text{ELSE IF } sWriteFwd[s] \neq Client$
$\qquad\qquad \text{THEN } BufSend(s, sWriteFwd[s], [op \mapsto m.op,$
$\qquad\qquad\qquad\qquad\qquad\qquad\qquad\qquad key \mapsto m.key,$
$\qquad\qquad\qquad\qquad\qquad\qquad\qquad\qquad val \mapsto m.val,$
$\qquad\qquad\qquad\qquad\qquad\qquad\qquad\qquad hops \mapsto Tail(m.hops)])$
$\qquad\qquad \text{ELSE UNCHANGED } \langle allBuf, allBufOp \rangle$

For a switch that is failed, forward the message to the forwarding address to simulate the failover and failure recovery logic
$SwitchFailedProc(s) \triangleq$
    $\wedge sBufSt[s] = \text{"busy"} \wedge sStatus[s] \neq \text{"alive"}$
    $\wedge sBufSt' = [sBufSt \text{ EXCEPT } ![s] = \text{"rdy"}]$
    $\wedge (SFailedRead(s) \vee SFailedWrite(s))$
    $\wedge \text{UNCHANGED } \langle sMem, sBuf, sStatus, sReadFwd, sWriteFwd, sFailedCount,$
                            $prevKVs, currentKVs, bufOpCount \rangle$

$SwitchProc(s) \triangleq SwitchAliveProc(s) \vee SwitchFailedProc(s)$

---

Failover and failure recovery of *NetChain* switches.

$SwitchMemCopy(old, new) \triangleq sMem' = [sMem \text{ EXCEPT } ![new] = sMem[old]]$

$SwitchBufClear(sSet) \triangleq$
    $\wedge sBuf' = [s \in Switches \mapsto$
                $\text{IF } s \in sSet \text{ THEN } [op \mapsto \text{"NoOp"}] \text{ ELSE } sBuf[s]]$
    $\wedge sBufSt' = [s \in Switches \mapsto$
                $\text{IF } s \in sSet \text{ THEN "rdy" ELSE } sBufSt[s]]$
    $\wedge allBuf' = [s \in AllNodes \mapsto$
              $\text{IF } s \in sSet$
              $\text{THEN } [t \in AllNodes \mapsto \langle \rangle]$
              $\text{ELSE } [t \in AllNodes \mapsto$
                      $\text{IF } t \in sSet \text{ THEN } \langle \rangle \text{ ELSE } allBuf[s][t]]]$
    $\wedge allBufOp' = [s \in AllNodes \mapsto$
              $\text{IF } s \in sSet$
              $\text{THEN } [t \in AllNodes \mapsto \text{FALSE}]$
              $\text{ELSE } [t \in AllNodes \mapsto$
                      $\text{IF } t \in sSet \text{ THEN FALSE ELSE } allBufOp[s][t]]]$

Take down a switch.
$SwitchFail(s) \triangleq$
    $\wedge sStatus[s] = \text{"alive"} \wedge s \in Range(SwitchChain)$
    $\wedge sStatus' = [sStatus \text{ EXCEPT } ![s] = \text{"failed"}]$
    $\wedge sFailedCount' = sFailedCount + 1$
    Read requests should be forwarded to previous hop in chain
    $\wedge sReadFwd' = [sReadFwd \text{ EXCEPT } ![s] = PrevInSeqOrX(s, Client)]$
    Write requests should be forwarded to next hop in chain
    $\wedge sWriteFwd' = [sWriteFwd \text{ EXCEPT } ![s] = NextInSeqOrX(s, Client)]$
    $\wedge SwitchBufClear(\{s\})$
    $\wedge \text{UNCHANGED } \langle sMem, prevKVs, currentKVs, bufOpCount \rangle$

Failure recovery: sync memory and update message forwarding address

$SwitchRecoveryTail(s) \triangleq$

$\quad\quad \wedge sStatus[s] =$ "failed"
$\quad\quad \wedge s = SwitchChain[Len(SwitchChain)]$
$\quad\quad \wedge (\text{LET } sNew \triangleq (\text{CHOOSE } x \in Switches :$
$\quad\quad\quad\quad\quad\quad\quad\quad x \notin Range(SwitchChain) \cup Range(sWriteFwd))$
$\quad\quad\quad\quad sPrev \triangleq PrevAliveInSeqOrClient[s]$
$\quad\quad \text{IN} \quad \wedge SwitchMemCopy(sPrev, sNew)$
$\quad\quad\quad\quad \wedge SwitchBufClear(\{sPrev, s\})$
$\quad\quad\quad\quad \wedge sStatus' = [sStatus \text{ EXCEPT } ![s] =$ "recovered"$]$
$\quad\quad\quad\quad \wedge sWriteFwd' = [sWriteFwd \text{ EXCEPT } ![s] = sNew]$
$\quad\quad\quad\quad \wedge sReadFwd' = [sReadFwd \text{ EXCEPT } ![s] = sNew])$
$\quad\quad \wedge \text{UNCHANGED } \langle sFailedCount, prevKVs, currentKVs, bufOpCount \rangle$

$SwitchRecoveryNotTail(s) \triangleq$
$\quad\quad \wedge sStatus[s] =$ "failed"
$\quad\quad \wedge s \in Range(SwitchChain)$
$\quad\quad \wedge s \neq SwitchChain[Len(SwitchChain)]$
$\quad\quad \wedge (\text{LET } sNew \triangleq (\text{CHOOSE } x \in Switches :$
$\quad\quad\quad\quad\quad\quad\quad\quad x \notin Range(SwitchChain) \cup Range(sWriteFwd))$
$\quad\quad\quad\quad sNext \triangleq NextAliveInSeqOrClient[s]$
$\quad\quad \text{IN} \quad \wedge SwitchMemCopy(sNext, sNew)$
$\quad\quad\quad\quad \wedge SwitchBufClear(\{sNext, s\})$
$\quad\quad\quad\quad \wedge sStatus' = [sStatus \text{ EXCEPT } ![s] =$ "recovered"$]$
$\quad\quad\quad\quad \wedge sWriteFwd' = [sWriteFwd \text{ EXCEPT } ![s] = sNew]$
$\quad\quad\quad\quad \wedge sReadFwd' = [sReadFwd \text{ EXCEPT } ![s] = sNew])$
$\quad\quad \wedge \text{UNCHANGED } \langle sFailedCount, prevKVs, currentKVs, bufOpCount \rangle$

$SwitchRecovery(s) \triangleq SwitchRecoveryTail(s) \vee SwitchRecoveryNotTail(s)$

---

Client related operators.

$CReadReq(k) \triangleq$
$\quad \text{LET} \quad ctail \triangleq SwitchChain[Len(SwitchChain)]$
$\quad\quad\quad\quad msg \triangleq [op \mapsto$ "Read"$, key \mapsto k, hops \mapsto \langle SwitchChain[Len(SwitchChain)] \rangle]$
$\quad \text{IN} \quad BufSend(Client, ctail, msg)$

$CWriteReq(k, v) \triangleq$
$\quad \text{LET } chead \triangleq SwitchChain[1]$
$\quad\quad\quad\quad msg \triangleq [op \mapsto$ "Write"$, key \mapsto k, val \mapsto \langle v, 0 \rangle, hops \mapsto SwitchChain]$
$\quad \text{IN} \quad BufSend(Client, chead, msg)$

Send a read (write) request with random key (key-value).
$ClientSend \triangleq \wedge \exists k \in Keys : CReadReq(k)$
$\quad\quad\quad\quad\quad\quad \vee (\exists v \in Values : CWriteReq(k, v))$
$\quad\quad\quad\quad \wedge \text{UNCHANGED } \langle sMem, sBuf, sBufSt, sStatus, sFailedCount,$
$\quad\quad\quad\quad\quad\quad\quad\quad\quad sReadFwd, sWriteFwd, prevKVs, currentKVs, bufOpCount \rangle$

Receive reply from *NetChain*, and update the observed values / versions accordingly.
$ClientRcv \triangleq \land \exists s \in Switches :$
$\qquad \land BufRcv(s, Client)$
$\qquad \land prevKVs' = currentKVs$
$\qquad \land currentKVs' = \text{LET } m \triangleq Head(allBuf[s][Client])$
$\qquad \qquad \qquad \qquad \qquad \text{IN } [currentKVs \text{ EXCEPT }![m.key] = m.val]$
$\qquad \land \text{UNCHANGED } \langle sMem, sBuf, sBufSt, sStatus, sReadFwd, sWriteFwd,$
$\qquad \qquad \qquad \qquad sFailedCount, bufOpCount \rangle$

---

Safety property: client can only observe key-values with incrementing version tags, even with switch failures and recovery.
$Consistency \triangleq \forall k \in Keys : prevKVs[k][2] \leq currentKVs[k][2]$

Update property: the version of a value in an upstream switch should be no less than the version of that in a downstream switch, or the switch is in failed status and not yet recovered.
$UpdatePropagation \triangleq \forall k \in Keys :$
$\qquad \forall s1 \in (Range(SwitchChain)) :$
$\qquad \forall s2 \in (Range(SwitchChain)) :$
$\qquad \quad SwitchToIdx(s1) < SwitchToIdx(s2) \Rightarrow$
$\qquad \quad \text{LET } up \triangleq \text{ IF } sStatus[s1] = \text{"alive"}$
$\qquad \qquad \qquad \qquad \text{THEN } s1$
$\qquad \qquad \qquad \qquad \text{ELSE IF } sStatus[s1] = \text{"recovered"}$
$\qquad \qquad \qquad \qquad \text{THEN } sWriteFwd[s1]$
$\qquad \qquad \qquad \qquad \text{ELSE } Client$
$\qquad \qquad down \triangleq \text{ IF } sStatus[s2] = \text{"alive"}$
$\qquad \qquad \qquad \qquad \text{THEN } s2$
$\qquad \qquad \qquad \qquad \text{ELSE IF } sStatus[s2] = \text{"recovered"}$
$\qquad \qquad \qquad \qquad \text{THEN } sWriteFwd[s2]$
$\qquad \qquad \qquad \qquad \text{ELSE } Client$
$\qquad \quad \text{IN} \quad \lor up = Client$
$\qquad \qquad \quad \lor down = Client$
$\qquad \qquad \quad \lor sMem[up][k][2] \geq sMem[down][k][2]$

---

Constraints to reduce state space explored by model checker.

$qConstraint \triangleq \forall src \in AllNodes :$
$\qquad \forall dst \in AllNodes :$
$\qquad \quad Len(allBuf[src][dst]) \leq maxQLen$

$sConstraint \triangleq sFailedCount \leq maxFailedCount$

$vConstraint \triangleq \forall k \in Keys : currentKVs[k][2] \leq maxVersion$

$bConstraint \triangleq bufOpCount \leq maxBufOpCount$

---

$Next \triangleq \lor ClientSend$
$\qquad\quad \lor ClientRcv$
$\qquad\quad \lor \exists\, s \in Switches : SwitchRcv(s)$
$\qquad\quad \lor \exists\, s \in Switches : SwitchProc(s)$
$\qquad\quad \lor \exists\, src \in Switches :$
$\qquad\qquad\quad \exists\, dst \in Switches :$
$\qquad\qquad\qquad BufOp(src, dst)$
$\qquad\quad \lor \exists\, s \in Switches : SwitchFail(s)$
$\qquad\quad \lor \exists\, s \in Switches : SwitchRecovery(s)$

$Spec \triangleq Init \land \Box[Next]\langle sMem,\ sBuf,\ sBufSt,\ sStatus,\ sReadFwd,\ sWriteFwd,$
$\qquad\qquad\qquad\quad sFailedCount,\ allBuf,\ allBufOp,$
$\qquad\qquad\qquad\quad prevKVs,\ currentKVs,\ bufOpCount\rangle$

---

THEOREM $Spec \Rightarrow \Box(TypeInvariant \land Consistency \land UpdatePropagation)$